\documentclass[letterpaper, 10 pt, conference]{ieeeconf}  

\IEEEoverridecommandlockouts                              

\usepackage{graphics} 
\usepackage{epsfig} 
\usepackage{mathptmx} 
\usepackage{times} 
\usepackage{amsmath} 
\usepackage{amssymb}  
\usepackage{subcaption}
\usepackage{gensymb}
\usepackage{xcolor}
\usepackage{cite}
\usepackage{algorithm}
\usepackage{algorithmic}
\usepackage{comment}
\title{\LARGE \bf
Peek into the Future Camera-based Occupant Sensing in Configurable Cabins for  Autonomous Vehicles}

\author{Avinash Prabu$^{1}$, Renran Tian$^{2}$, Lingxi Li$^{1}$, Jialiang Le$^{3}$, Srinivasan Sundararajan$^{3}$, and Saeed Barbat$^{3}$
\thanks{The authors would thank Ford to support the research work.}%
\thanks{$^{1}$Avinash Prabu and Lingxi Li are with the Department of Electrical \& Computer Engineering,
        Indiana University Purdue University Indianapolis, 723 West Michigan Street, Indianapolis, IN, USA
        {\tt\small avinash.prabu21@gmail.com, LL7@iu.edu}}%
\thanks{$^{2}$Renran Tian is with the Department of Computer Information Technology, Indiana University Purdue University Indianapolis, 799 West Michigan Street, Indianapolis, IN, USA
        {\tt\small rtian@iupui.edu}}%
\thanks{$^{3}$Jerry Le, Srinivasan Sundararajan and Saeed Barbat are with the Ford Motor company,
        Research and Innovation Center, Dearborn, MI, USA. 
        {\tt\small jle1@ford.com, ssundar1@ford.com, sbarbat@ford.com,}}%
}

\begin{document}

\maketitle
\thispagestyle{empty}
\pagestyle{empty}

\begin{abstract}
The development of fully autonomous vehicles (AVs) can potentially eliminate drivers and introduce unprecedented seating design. However, highly flexible seat  configurations may lead to occupants' unconventional poses and actions. Understanding occupant behaviors and prioritize safety features become eye-catching topics in the AV research frontier. Visual sensors have the advantages of cost-efficiency and high-fidelity imaging and become more widely applied for in-car sensing purposes. Occlusion is one big concern for this type of system in crowded car cabins. It is important but largely unknown about how a visual-sensing framework will look like to support 2-D and 3-D human pose tracking towards highly configurable seats. As one of the first studies to touch this topic, we peek into the future camera-based sensing framework via a simulation experiment.  Constructed representative car-cabin, seat layouts, and occupant sizes, camera coverage from different angles and positions is simulated and calculated. The comprehensive coverage data are  synthesized through an optimization process to determine the camera layout and overall occupant coverage. The results show the needs and design of a different number of cameras to fully or partially cover all the occupants with changeable configurations of up to six seats.
\end{abstract}

\section{INTRODUCTION}

The development of autonomous vehicles (AVs) has been rapidly growing, leading to more machine control and lesser human interventions. This revolutionary progress will eventually remove drivers from the cars and give occupants the freedom to carry out more flexible activities. Besides the expected improvement of mobility, efficiency, and comfort, this change also increases the difficulty of efficiently triggering safety features and better protecting occupants when crashes happen. Without driver interventions, AVs need to prioritize safety features based on external conditions and occupant behaviors via enhanced sensor systems \cite{USDOT2020}. Thus, "researching occupant protection in alternative vehicle designs" becomes a higher-focused research topic by the National Science and Technology Council and the United States Department of Transportation.  

\subsection{Related Research}

Towards traditional manual-driving vehicles, driver monitoring systems have been frequently researched. Different impaired driver states like distraction \cite{yekhshatyan2012changes, tian2013studying}, drowsiness \cite{sahayadhas2012detecting}, fatigue \cite{lal2001critical,mollicone2019predicting}, and mind wandering \cite{baldwin2017detecting} are widely investigated about their effects on driving safety. Research also focuses on the corresponding detection algorithms and mitigation means. Although physiological sensors like EEG, ECG, heart rate, and other measurements are valid and reliable indicators for different driver states \cite{baldwin2017detecting,sahayadhas2012detecting,gao2019eeg}, camera-based systems are more widely developed and implemented.  Studies and practices show that visual sensors are non-intrusive solutions and more convenient to be integrated into vehicle-cabins \cite{yekhshatyan2012changes,horng2004driver,sahayadhas2012detecting, wang2016driver}. These driver state sensing and mitigation systems are initially designed to reduce impaired driver status and improve safety. However, with the advantage of vehicle automation levels, driver state sensing also extends its functionalities and is used in more complicated conditions.  

As level 2 and level 3 autonomous vehicles are coming into reality, driver monitoring systems become core components of the overall advanced driver assist systems and the autonomous driving systems. AVs road test in California has demonstrated the problems of automation disengagement and highlighted the importance of the transition process from auto-control to manual-control \cite{favaro2018autonomous, dixit2016autonomous}. This control hand-over process has a list of human factors concerns, including driver inattention and distraction, situational awareness, and over-reliance on the technology \cite{cunningham2015autonomous}. These issues may result in a take-over reaction time between 2 to 10 seconds, a prolonged take-over completion time from 2 to 20 seconds \cite{eriksson2017takeover}, and even longer time for a driver to fully stabilize the car \cite{merat2014transition}. Considering the various take-over scenarios in terms of urgency, predictability, and criticality \cite{gold2017testing}, it is important to monitor the driver's states before and during the take-over process to re-engage them via optimized interface design \cite{cunningham2015autonomous}, and adopt appropriate transition strategies \cite{lu2016human}. 

Although more attention starts to focus on rear-seat occupants' safety, full-cabin occupant sensing is much less studied than the driver state sensing systems. Not surprisingly, rear-seat occupants have more differences in terms of individual characteristics \cite{durbin2015rear} when these seats are treated as safer for young and older passengers and people in a more vulnerable status. However, these rear-seat occupants are equally likely to be involved in car crashes \cite{durbin2015rear}, and their protection becomes a more challenging task. Most of the full-cabin or rear-seat occupant sensing studies implement seat-embedded sensors to detect vital signs or occupancy of different seats, with one typical study exampled in \cite{orlewski2011advanced}. In recent years, researchers started to use seat-independent sensors like Kinect \cite{loeb2017automated} and Ultra Wide-Band Radar \cite{deng2019efficient} to detect and track all occupants, including those who sit in the back. However, most of these studies are still in the initial investigation states and usually only focus on traditional fixed cabin layouts.

\subsection{Research Objectives and Contributions}
To our best knowledge, few studies discuss the sensing of all occupants in a highly configurable vehicle cabin. 
To address this research gap, this study focuses on camera-based full-cabin sensing given such a dynamic environment. A car cabin was created on a simulation platform with multiple seat layouts. Occupants/passengers were designed, and their key upper body points were identified. Computer simulation was first conducted to obtain the camera coverage of the key points across all occupants in different seat layouts from different angles and positions. An algorithm was then developed to find the combinations of camera locations and angles to optimize overall coverage towards different goals. The contributions of the paper are itemized below: 
\begin{itemize}
    \item Propose the concept of full-cabin occupant sensing with a highly configurable seat layouts. 
    \item Describe the possible designs of camera-based sensing framework for L3+ autonomous vehicles. 
    \item Provide a simulation platform to develop and evaluate camera-based occupant sensing systems. 
    \item Create a novel optimization strategy to study camera locations and orientations, given the constraints on the coverage requirements and camera numbers.
\end{itemize}


\begin{figure}[H]
    \centering
\begin{subfigure}{0.9\linewidth}
\includegraphics [width=\linewidth]{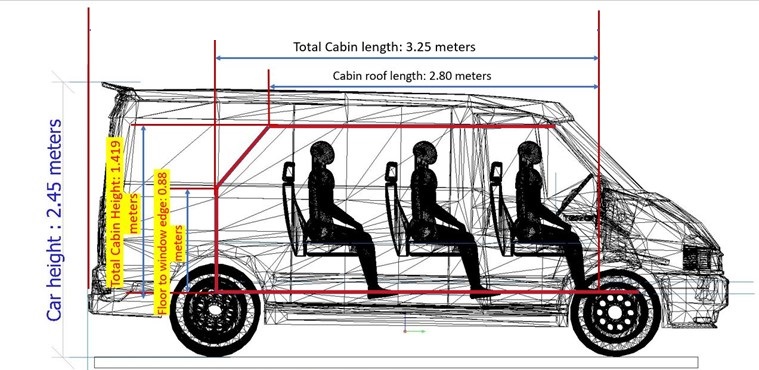}
\caption{Side View}
\label{fig:sideview}
\end{subfigure}
~
\begin{subfigure}{0.5\linewidth}
\includegraphics [width=\linewidth]{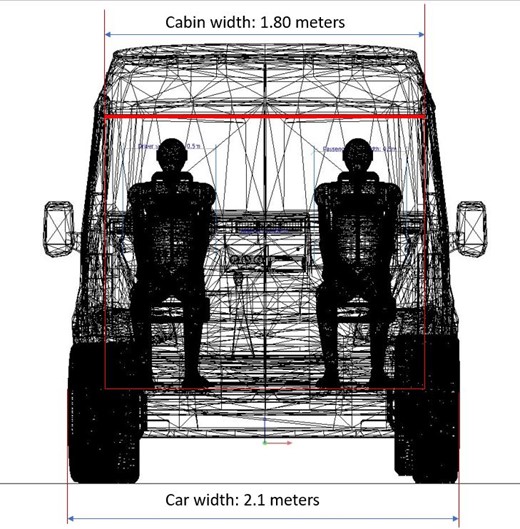}
\caption{Back View}
\label{fig:backview}
\end{subfigure}
\caption{Cabin Design}
\label{fig:cabindesign}
\end{figure}

\section{Simulation Method}

\subsection{Cabin Design}

In order to set up a representative cabin, we surveyed a list of on-market vehicles, including SUVs and mini-vans to acquire average cabin measurements. The results of the survey is explained in Table \ref{Ta:cabinDesign}.

\begin{table}[ht]
\caption{Cabin Measurements}
\begin{center}

\begin{tabular}{|c|c|c|}
\hline
Axis & Parameter                & Measurement (m) \\\hline
Z    & Cabin height             & 1.41            \\
     & Occupant head to ceiling    & 0.11            \\\hline
Y    & Cabin length       & 3.25            \\
     & Leg room row 1     & 0.11            \\
     & Leg room row 2     & 0.5             \\
     & Leg room row 3     & 0.35            \\\hline
X    & Cabin width              & 1.8             \\
     & Driver seat from side    & 0.15            \\
     & Passenger seat from side & 0.15         \\\hline 
\end{tabular}
\end{center}
\label{Ta:cabinDesign}
\end{table}

The angle of the rear window is also averaging existing car cabins. The final design of the cabin is shown in Fig. \ref{fig:cabindesign}. The red line outlines the cabin inside a Ford Transit van from two angles to illustrate the designed cabin sizes. The cabin shape is not based on any existing car but represents a typical size of current three-row vehicles. The cabin design sets up the simulation boundaries. The same method presented in this study can be applied accordingly by changing the cabin dimensions for any given car. 

\subsection{Key Body Points}
\begin{figure}[h!]
    \centering
    \centering
    \includegraphics [scale=0.5] {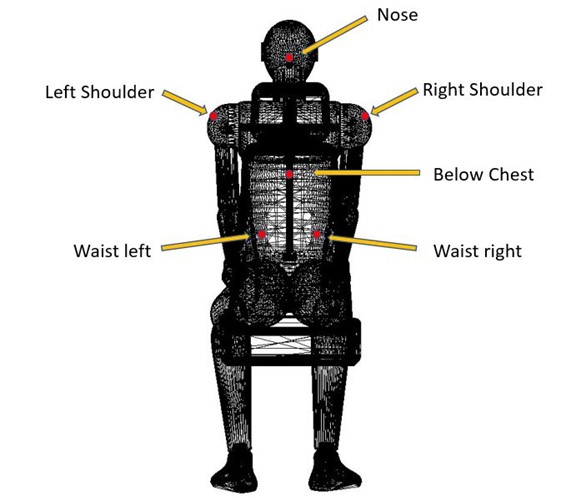}
    \caption{Key Body Points}
    \label{fig:markerplacement}
\end{figure}

In the simulation environment, we focus on body key points for each seated occupant. The simulation process will output the camera coverage of each of these key points for each occupant in all interested seat layouts from different positions and angles in the cabin. 
This research defines six body locations, including nose, left shoulder, right shoulder, chest, right waist, and left waist. The body locations are depicted in Fig. \ref{fig:markerplacement}. 

\label{Sec:MarkerPlacement}

\subsection{Simulation Space}
\begin{figure}[h!]
    \centering
    \centering
    \includegraphics [scale=0.45] {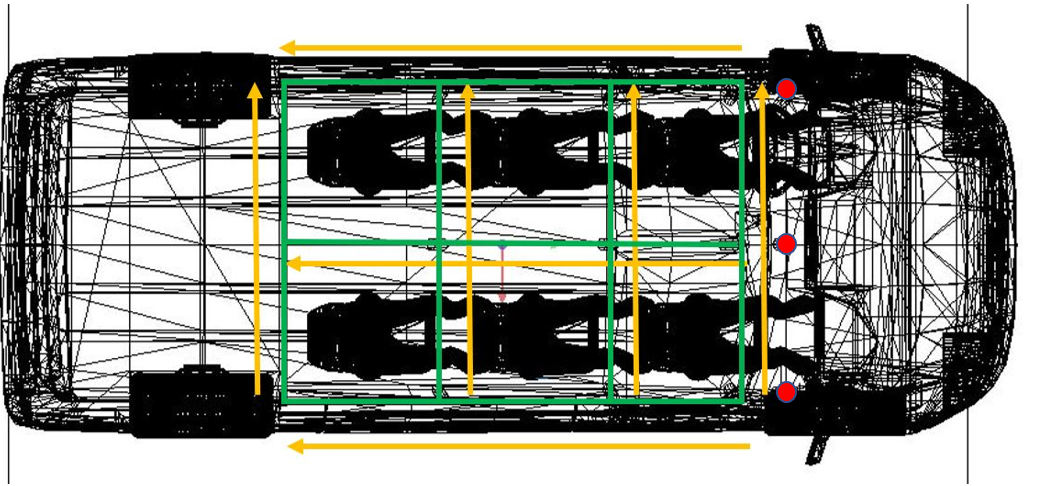}
    \caption{Camera Locations}
    \label{fig:camloc}
\end{figure}

In the simulation environment, we move the camera along a constructed frame in the cabin. The paths along which the camera moves is shown as green lines in Fig. (\ref{fig:camloc}), following a step size of 0.1m. At each step, the cameras are rotated at various roll, pitch, and yaw angles. At all the locations and angles, the coverage for each body key point is recorded. Three camera locations on the windshield (illustrated as red dots) are also recorded. 

\label{Sec:SimulationSpace}

\subsection{Simulation Outputs}
\begin{figure*}[ht!]
    \centering
    \centering
    \includegraphics [scale=0.7] {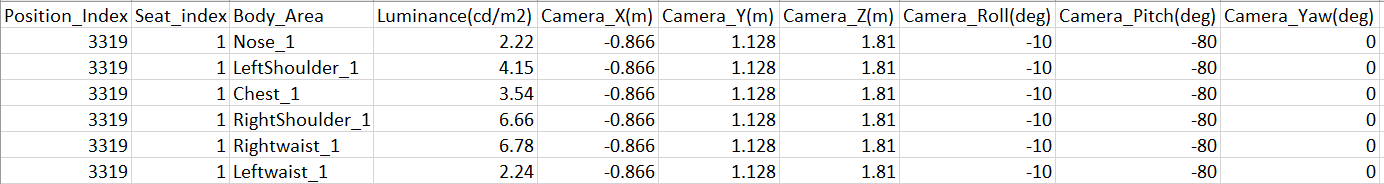}
    \caption{Example of Dataset}
    \label{fig:dataset}
\end{figure*}

The results of the simulation are exported as a .csv file. The variables in the dataset are \emph{Position\_Index, Seat\_Index, Body\_Area, Coverage Indicator, Camera\_X, Camera\_Y, Camera\_Z, Camera\_Roll, Camera\_Pitch, and Camera\_Yaw}. The \emph{Position\_Index} is a unique ID to represent the position of the camera. The \emph{Seat\_index} represents the seat for the occupant in the vehicle. \emph{Body\_Area} is the body key point, as explained in \ref{Sec:MarkerPlacement}. The \emph{Coverage Indicator/Luminance} is a number showing coverage of the key point if the value is non-zero (in this research, we use lighting sources to replace cameras, and use light dominance on different body parts to estimate camera coverage as well as distance). The \emph{Camera\_X, Camera\_Y, Camera\_Z, Camera\_Roll, Camera\_Pitch, Camera\_Yaw} provides the pose of the camera at a particular position, based on the vehicle coordinate system. An example of the data is provided in Fig. (\ref{fig:dataset}).
\subsection{Simulation Setup}
The simulation aims to find the coverage towards each occupant in multiple seat layouts from all possible angles and positions of the cameras in the cabin. 
As an equivalent process, we decide to use light coverage to represent camera coverage to use convenient off-the-shelf software. The idea behind this is that in a 3D environment, if light can illustrate a body part in the cabin, then a camera in the same position/angles should also capture the body part. The light can be adjusted with the same angular ranges as the camera and can be blocked by the same obstacles.

With that, a simulation environment is firstly constructed in a 3D virtual environment, including a car CAD model and movable objects like seats, occupants, and lighting sources. There are several main features of the simulation environment:
\begin{enumerate}
    \item Any car CAD model can be imported into the simulation environment to reflect the actual cabin dimensions.
    \item The movable objects can be arranged into different design layouts in the cabin.
    \item The angles of the lights can be adjusted to represent the simulated camera's view angles. In this study, we set up the camera diagonal view range as 94 degrees. 
    \item Multiple virtual sensors can be attached to the occupants at desired body locations, which will report the strength of light coverage.
\end{enumerate}

Based on sensor outputs, we can determine if certain body parts (where the sensors are attached) are visible to the camera set up at the same locations/angles as the lights. A program will automatically change the position and orientation values for the light and enumerate all possible angles and positions of interest. During this automated process, one light will move along predetermined routes on the ceiling of the cabin and the windshield window at a step of 10cm. At each light location, the light will rotate Roll, Pitch, Yaw angles in steps of 10 degrees. At each location (XYZ coordinates) and each direction (RPY angles) of the light, the script will record all sensors' readings attached to all occupants. These readings directly measure the coverage of corresponding body parts from a camera in the same pose. 

This process can be completed for all the interested seat layouts separately, and the results can then be integrated. As these light simulation data are reflecting the coverage of cameras, an optimization algorithm (explained in Section \ref{opt}) will select the optimal combinations of different camera locations and angles to maximize the coverage of all occupants in all simulated seat-layouts.

\section{Optimization Algorithm}
\label{opt}
The main goal of the optimization algorithm is to cover all the \emph{Body\_Area} of all the occupants with minimum number of camera. The optimization uses the dataset from the simulation to find the best possible positions of the camera. The simulation provides the intensity of light (\emph{Luminance}) on each marker as one of the primary variables. As explained earlier, if the light falls on the marker (\emph{Luminance}$>$0), then the camera can capture the marker and if the marker is not illuminated (\emph{Luminance}=0), then the camera cannot capture the marker. With this idea in mind, the first step in the algorithm is to convert the luminance measurements into binary values. This is explained in Algorithm \ref{algo:BLUM}. 

\begin{algorithm}

\caption{Convert Luminance to Binary Luminance} \label{algo:BLUM}
\begin{algorithmic}
\STATE $n \rightarrow$ total number of rows in the dataset  
\STATE $BLuminance=[0]*len(Luminance)$
\FOR{$i=1:n$}
\IF{$Luminance(i)>0$}
\STATE $BLuminance(i)=1$
\ENDIF
\IF{$Luminance(i)=0$}
\STATE $BLuminance(i)=0$
\ENDIF
\ENDFOR
\end{algorithmic}
\end{algorithm}

\begin{equation}
\label{eqn:luminance}
    \begin{aligned}
    Luminance &= [L1 & L2 && L3 ]\\
    L1 & = [1.2 & 5.6 && 3.4 && 9.3 && 6.5 && 0 && 0 && 0 \\
    & & 0 && 0 && 0 && 0] \\
    L2 & = [0 & 0 && 0 && 0 && 0 && 0 && 0 && 6.8 \\
    & & 4.3 && 2.2 && 8.7 && 9.1] \\
    L3 & = [0 & 0 && 0 && 0 && 1.5 && 4.9 && 11.2 && 3.5 \\
    & & 0 && 0 && 0 && 0] 
    \end{aligned}
\end{equation}
 For example, assuming that Equation. (\ref{eqn:luminance}) provides the luminance of a two occupant seat layout for three camera positions. After running through Algorithm \ref{algo:BLUM}, the binary luminance is obtained as given in Equation (\ref{eqn:Bluminance}).
\begin{equation}
\label{eqn:Bluminance}
    \begin{aligned}
    BLuminance &= [BL1 & BL2 && BL3 ]\\
    BL1 & = [1 & 1 && 1 && 1 && 1 && 0 && 0 && 0 \\
    & & 0 && 0 && 0 && 0] \\
    BL2 & = [0 & 0 && 0 && 0 && 0 && 0 && 0 && 1 \\
    & & 1 && 1 && 1 && 1] \\
    BL3 & = [0 & 0 && 0 && 0 && 1 && 1 && 1 && 1 \\
    & & 0 && 0 && 0 && 0] 
    \end{aligned}
\end{equation}

The next step in the algorithm is to arrange the \emph{Bluminance} in matrix form. This step helps in moving forward with formulating the optimization problem. Each column of the matrix represents a particular position's Bluminance.  Assuming there are $m$ total positions that were simulated and there are $p$ occupants in the scenario, the dimension of the matrix will be $6p \times m$. As mentioned earlier, there are six markers for each occupant and each marker has a luminance value, which equates to $6p$ \emph{Bluminance} values for each position. The $q^{th}$ column of the \emph{BMatrix} corresponds to the $q^{th}$ position.

\begin{algorithm}
\caption{Arrange $BLuminance$ in matrix form} \label{algo:BMAT}
\begin{algorithmic}
\STATE $m \rightarrow$ total number of positions in the dataset  
\STATE $p \rightarrow$ number of occupants in the scenario
\STATE $BMatrix \rightarrow$ zero matrix of dimension $6p$ x $m$
\STATE $j=1$
\FOR{$i=1:m$}
\STATE $BMatrix[:,i] = BLuminance[j:j+6p-1]$
\STATE $j=j+6p$
\ENDFOR
\end{algorithmic}
\end{algorithm}

Continuing with the example provided in Equation. (\ref{eqn:Bluminance}), the BMatrix can be obtained using Algorithm \ref{algo:BMAT}. The BMatrix is given in Equation. (\ref{eqn:BMAT}). 

\begin{equation}
BMatrix=
\begin{bmatrix}
1 & 0 & 0 \\
1 & 0 & 0 \\
1 & 0 & 0 \\
1 & 0 & 0 \\
1 & 0 & 1 \\
0 & 0 & 1 \\
0 & 0 & 1 \\
0 & 1 & 1 \\
0 & 1 & 0 \\
0 & 1 & 0 \\
0 & 1 & 0 \\
0 & 1 & 0 
\end{bmatrix}
\label{eqn:BMAT}
\end{equation}

The optimization proceeds with the \emph{BMatrix} as one of the primary parameters. The main goal of the optimization is to minimize the number of cameras and to make sure all the markers are covered. The standard form of the optimization problem is provided in Equation (\ref{eqn:opt1}). 

\begin{equation}
\label{eqn:opt1}
    \begin{aligned}
    & \underset{x}{\text{min}}
    & & c^Tx \\
    & \text{s.t}
    & & Ax \geq b, \;  \\
    &&& x=\{0,1\} \\
    \end{aligned}
\end{equation}

In Equation (\ref{eqn:opt1}), the one dimensional vector $c$ has all its elements as 1. The dimension of $c$ is $m \times 1$. The vector $x$ is the solution to the equation $Ax \geq b$, where $A$ is the $BMatrix$ and $b$ is a vector with all the elements as 1, with dimension $6p\times1$. The vector $b$ indicates that all the \emph{Body\_Area} needs to be covered. This ensures that the objective function is the sum of all the elements of $x$. It can be tweaked to accommodate other requirements also. For obvious reasons, the constraint on elements of $x$ is that it can be only 0 or 1. Solving for $x$ using the above optimization problem will provide the best position to place the cameras in the cabin. The vector $x$ will have 1's in the indices of the required positions. 

With the $BMatrix$ obtained in the example given in Equation. (\ref{eqn:BMAT}), the only possible solution to the optimization problem is the column vector $x=[1 1 1]^T$. The value $Ax$ is given in Equation. (\ref{eqn:Ax}) and the value of $c^Tx$ is 3. 

\begin{equation}
\label{eqn:Ax}
    \begin{aligned}
   Ax & = [1 & 1 && 1 && 1 && 2 && 1 && 1 && 2 \\
    & & 1 && 1 && 1 && 1] \\
    \end{aligned}
\end{equation}

The optimization problem can also be modified to accommodate various other constraints. Assuming that the user can afford only two cameras in the cabin and requires the algorithm to cover maximum possible markers. In this case, the optimization algorithm can be set as given in Equation (\ref{eqn:opt2}).

\begin{equation}
\label{eqn:opt2}
    \begin{aligned}
    & \underset{x,b}{\text{max}}
    & & \lambda^Tb \\
    & \text{s.t}
    & & Ax = b, \;  \\
     &&& c^Tx=2, \; \\
    &&& x=\{0,1\} \; \\
    &&& b=\{0,1\} 
    \end{aligned}
\end{equation}

In Equation (\ref{eqn:opt2}), $\lambda$ is a one dimensional vector with all the elements as 1 and the dimension is $6p\times 1$. $b$ is a one dimensional vector with dimension $6p\times1$, which indicates the coverage of markers.  The objective function ensures the sum of all the elements of $b$ is a global maximum across all possible combinations of two cameras. The one dimensional vector $c$ has all its elements as 1. The dimension of $c$ is $m\times1$. The vector $x$ is consistent with the explanation given for Equation (\ref{eqn:opt1}), which stands for the locations of the cameras needed. $c^Tx =2$ ensures that only a total of two cameras can be used. The constraints on $x$ and $b$ is to make sure that the results don not contain cameras at the same location. Since this optimization problem cannot be solved with linear programming, we use an iterative method to find all the possible combinations that will provide the best coverage. 

Again, continuing from the example given in Equation. (\ref{eqn:BMAT}), if the constraint on number of cameras is 2, the best coverage possible with two cameras is given in Equation. (\ref{eqn:b}), and the best camera positions to obtain this coverage is $x=[1 1 0]^T$. 

\begin{equation}
\label{eqn:b}
    \begin{aligned}
  Ax = b & = [1 & 1 && 1 && 1 && 1 && 0 && 0 && 1 \\
    & & 1 && 1 && 1 && 1] \\
    \end{aligned}
\end{equation}

\section{Optimization Result and Visualization}
Three experiments are conducted using the developed simulation and optimization methods to picture how a future camera-based full-cabin occupant sensing frame may look like and perform in highly configurable cabins. Results are reported and illustrated separately in this section. 

\subsection{Experiment 1: Coverage of all occupants for six seat layouts independently}

\begin{figure}[h!]
\centering
    \includegraphics [scale=0.5] {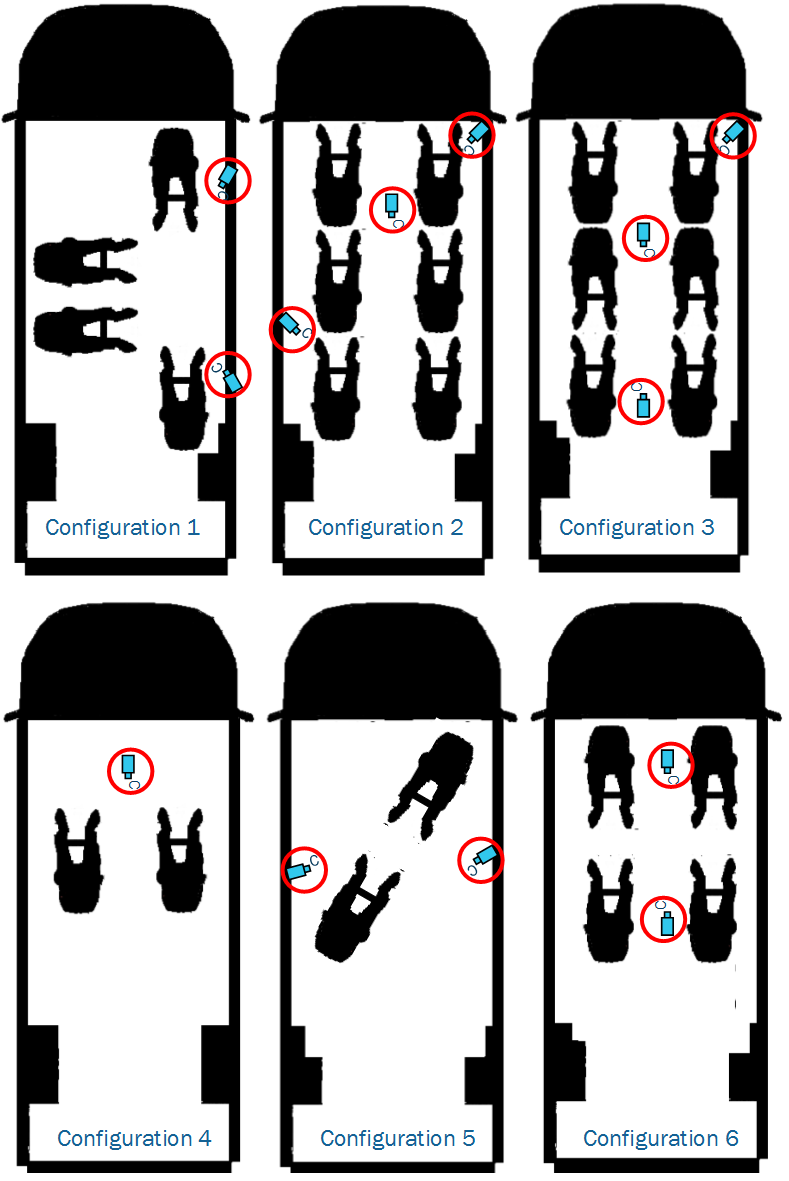}
\caption{Optimized Camera Positions}
\label{fig:OptViz}
\end{figure}

The simulation was performed on the six configurations given in Fig. \ref{fig:OptViz}. To better demonstrate all possibilities, these configurations have two (configurations 4 and 5), four (configurations 1 and 6), and six (configurations 2 and 3) occupants correspondingly. 

The optimization results show that configurations 2 and 3 require a minimum of three cameras to cover all the markers. Configurations 1, 5 and 6 require two cameras to cover all the occupants and configuration 4 requires just one camera. The results of the optimized positions are shown in Fig. \ref{fig:OptViz}. In Fig. \ref{fig:OptViz}, the red circles represent the camera positions. 

\begin{figure}[h!]
    \centering
\begin{subfigure}{0.3\linewidth}
\includegraphics [width=\linewidth]{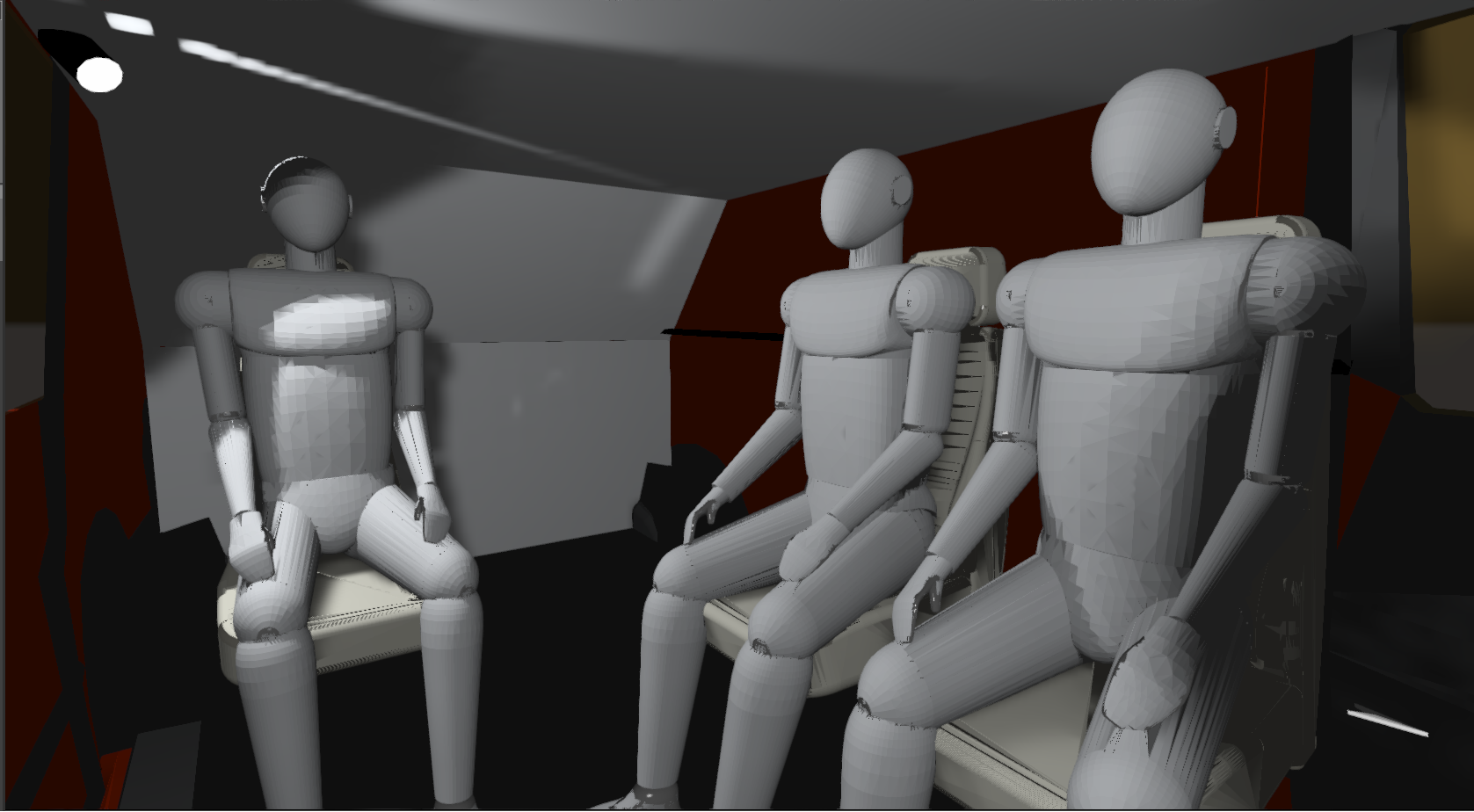}
\caption{Layout 1-1}
\label{fig:C1-1}
\end{subfigure}
~
\begin{subfigure}{0.3\linewidth}
\includegraphics [width=\linewidth]{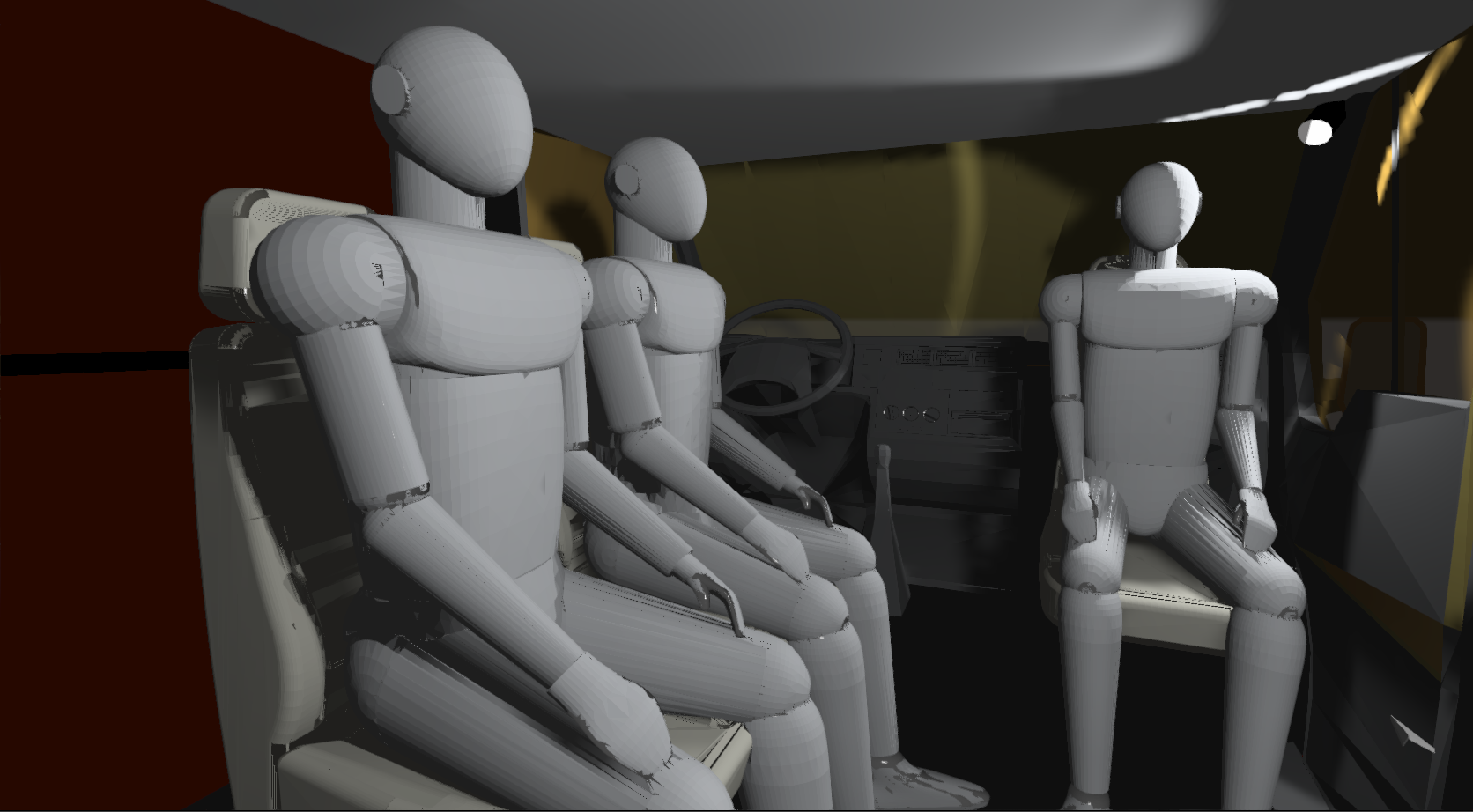}
\caption{Layout 1-2}
\label{fig:C1-2}
\end{subfigure}
~
\begin{subfigure}{0.3\linewidth}
\includegraphics [width=\linewidth]{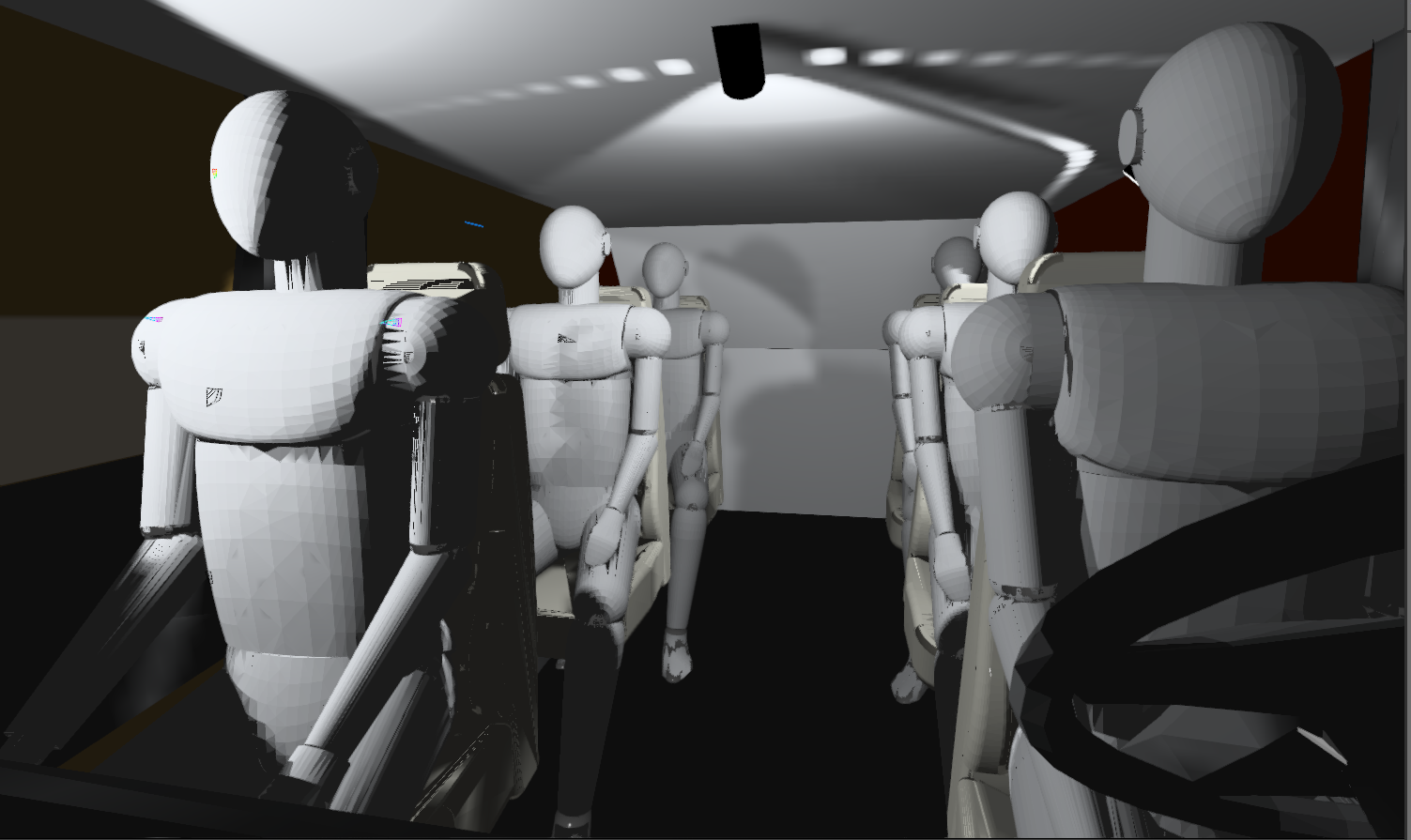}
\caption{Layout 2}
\label{fig:C2}
\end{subfigure}
~
\begin{subfigure}{0.3\linewidth}
\includegraphics [width=\linewidth]{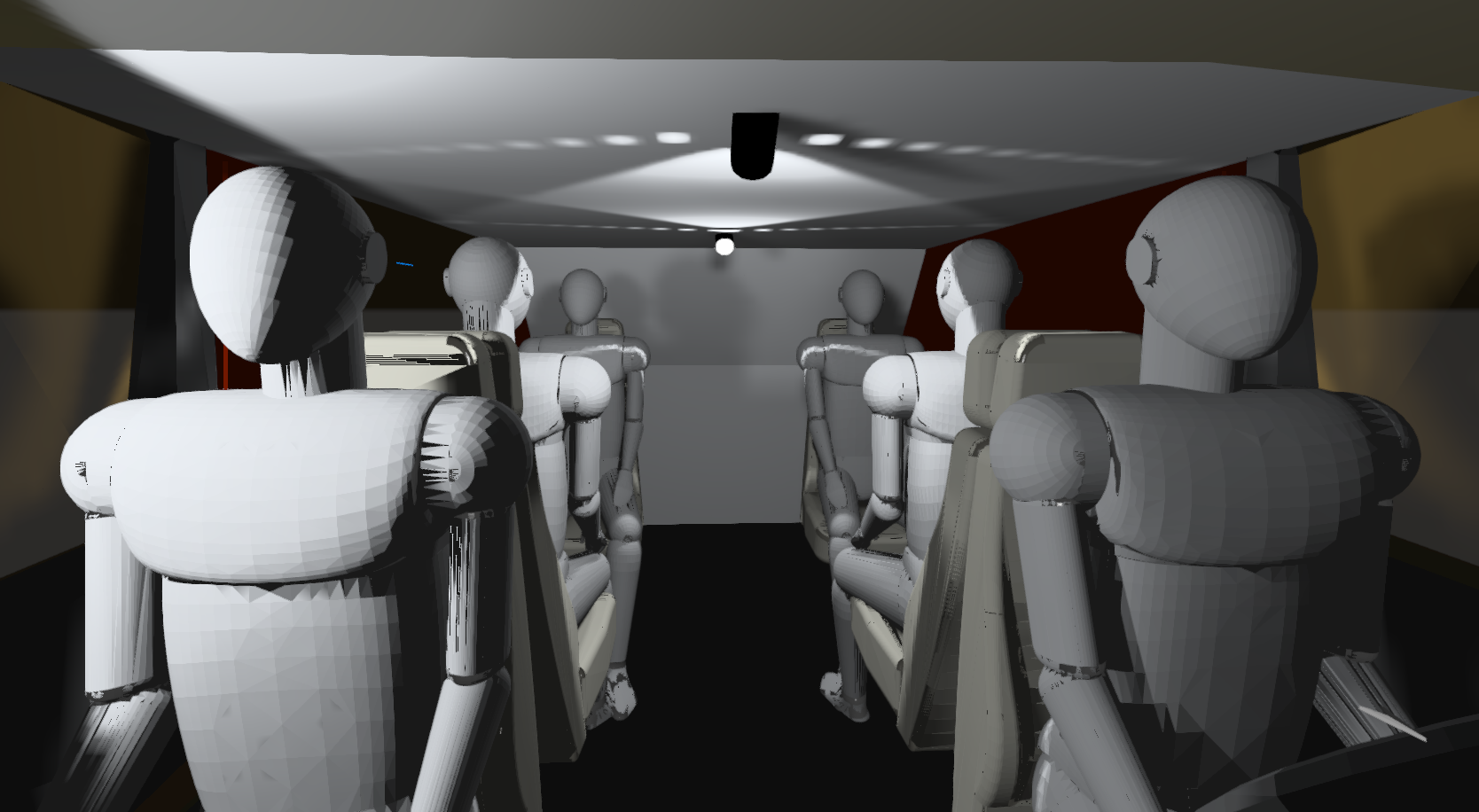}
\caption{Layout 3-1}
\label{fig:C3-1}
\end{subfigure}
~
\begin{subfigure}{0.3\linewidth}
\includegraphics [width=\linewidth]{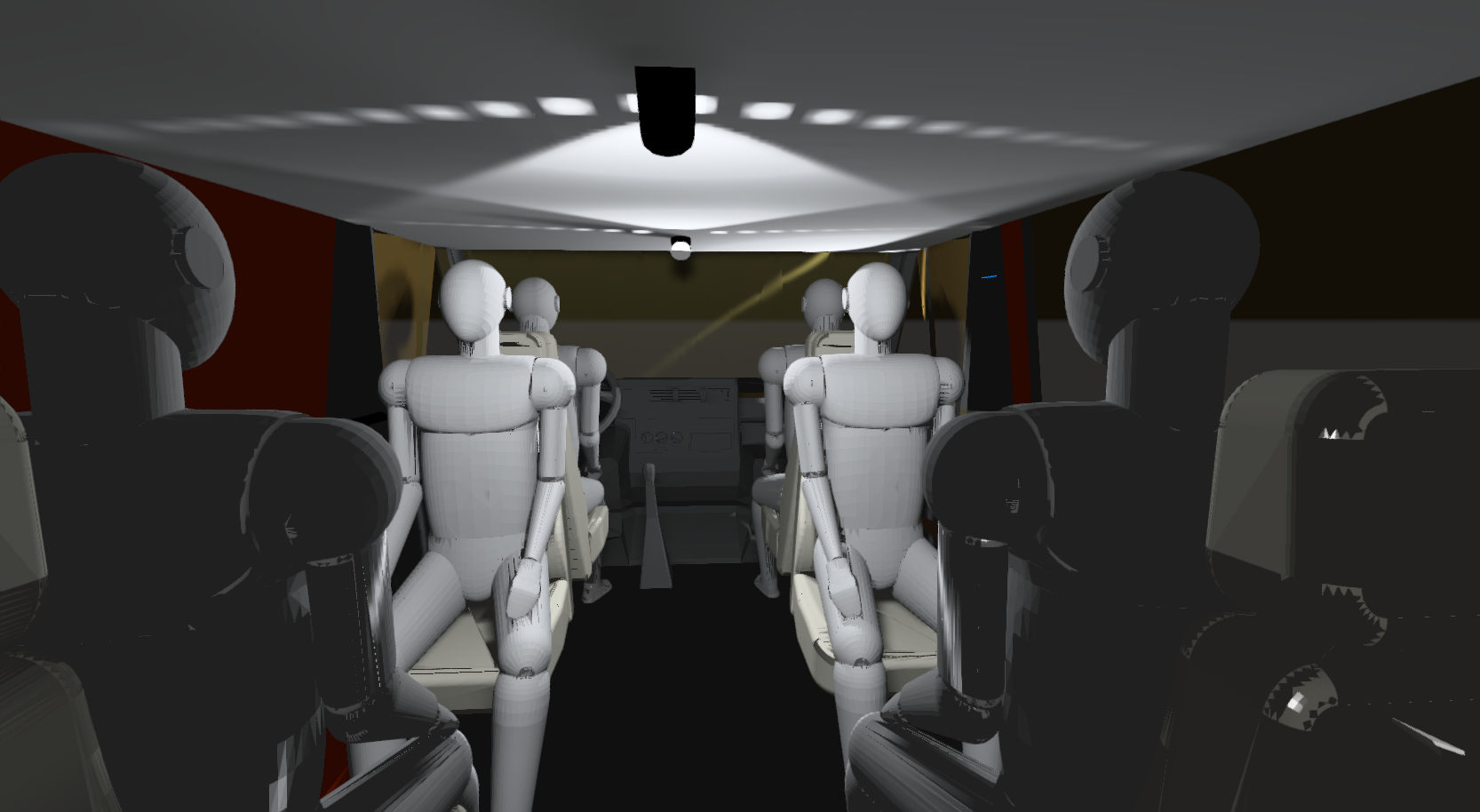}
\caption{Layout 3-2}
\label{fig:C3-2}
\end{subfigure}
~
\begin{subfigure}{0.3\linewidth}
\includegraphics [width=\linewidth]{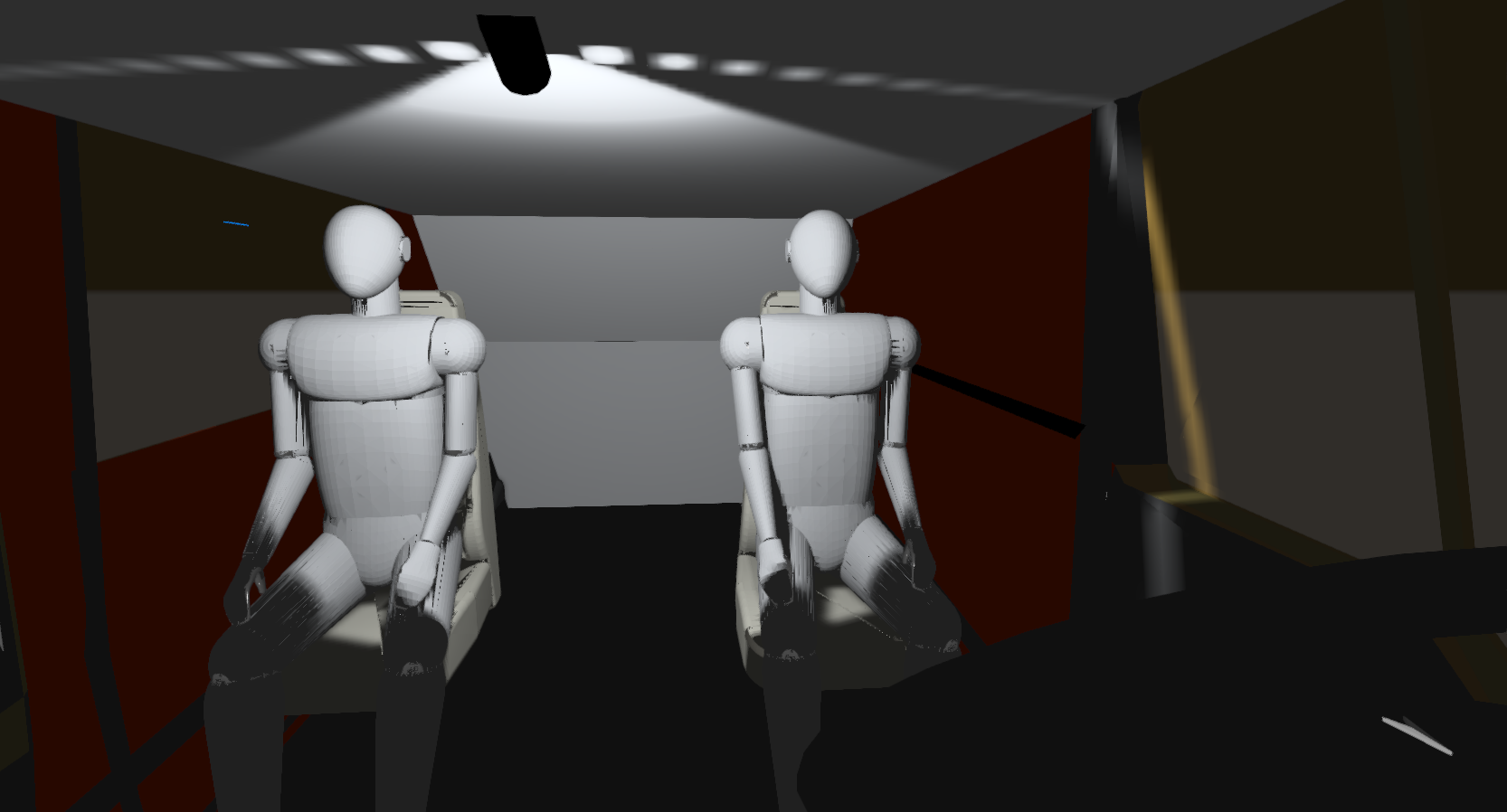}
\caption{Layout 4}
\label{fig:C4}
\end{subfigure}
~
\begin{subfigure}{0.3\linewidth}
\includegraphics [width=\linewidth]{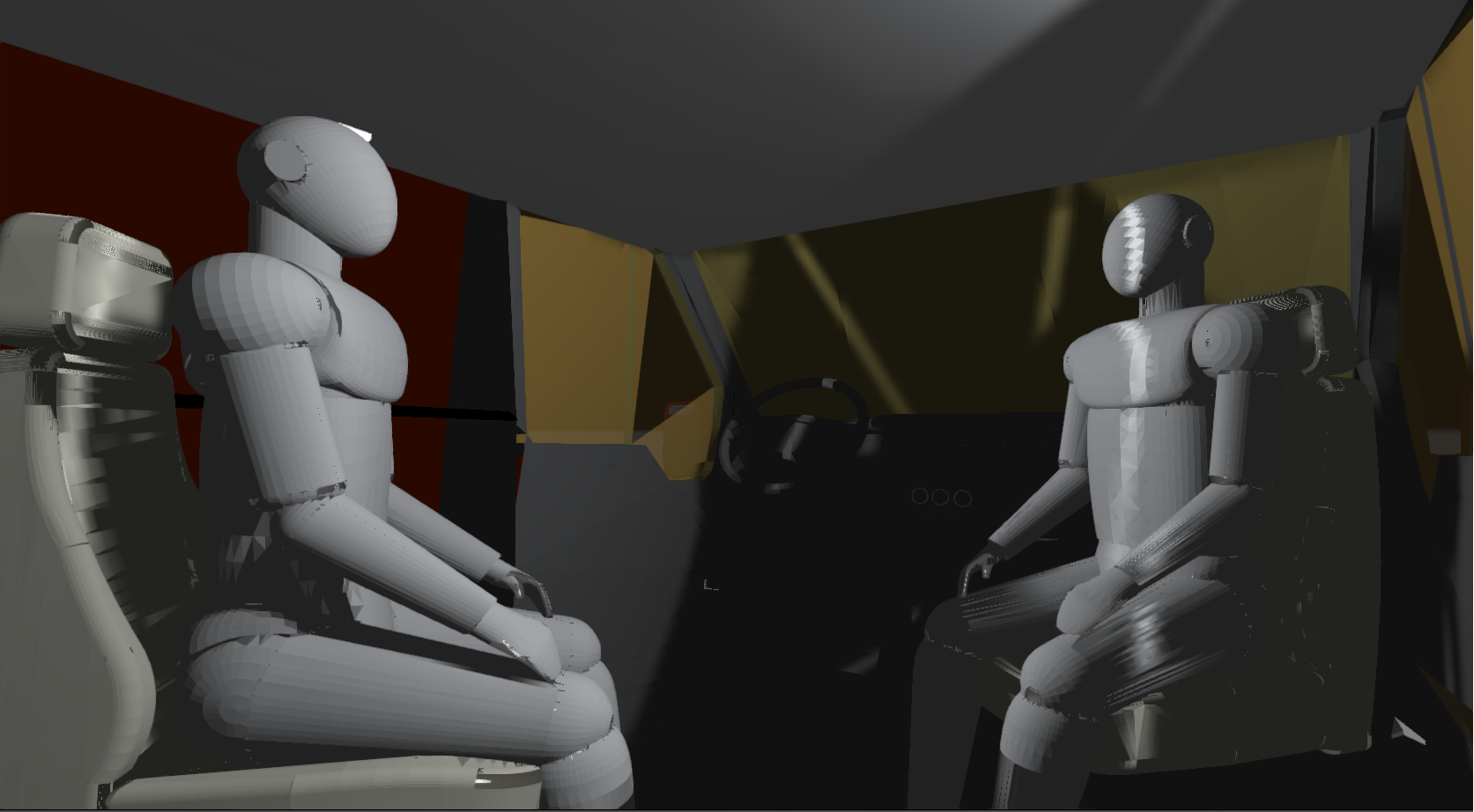}
\caption{Layout 5}
\label{fig:C5}
\end{subfigure}
~
\begin{subfigure}{0.3\linewidth}
\includegraphics [width=\linewidth]{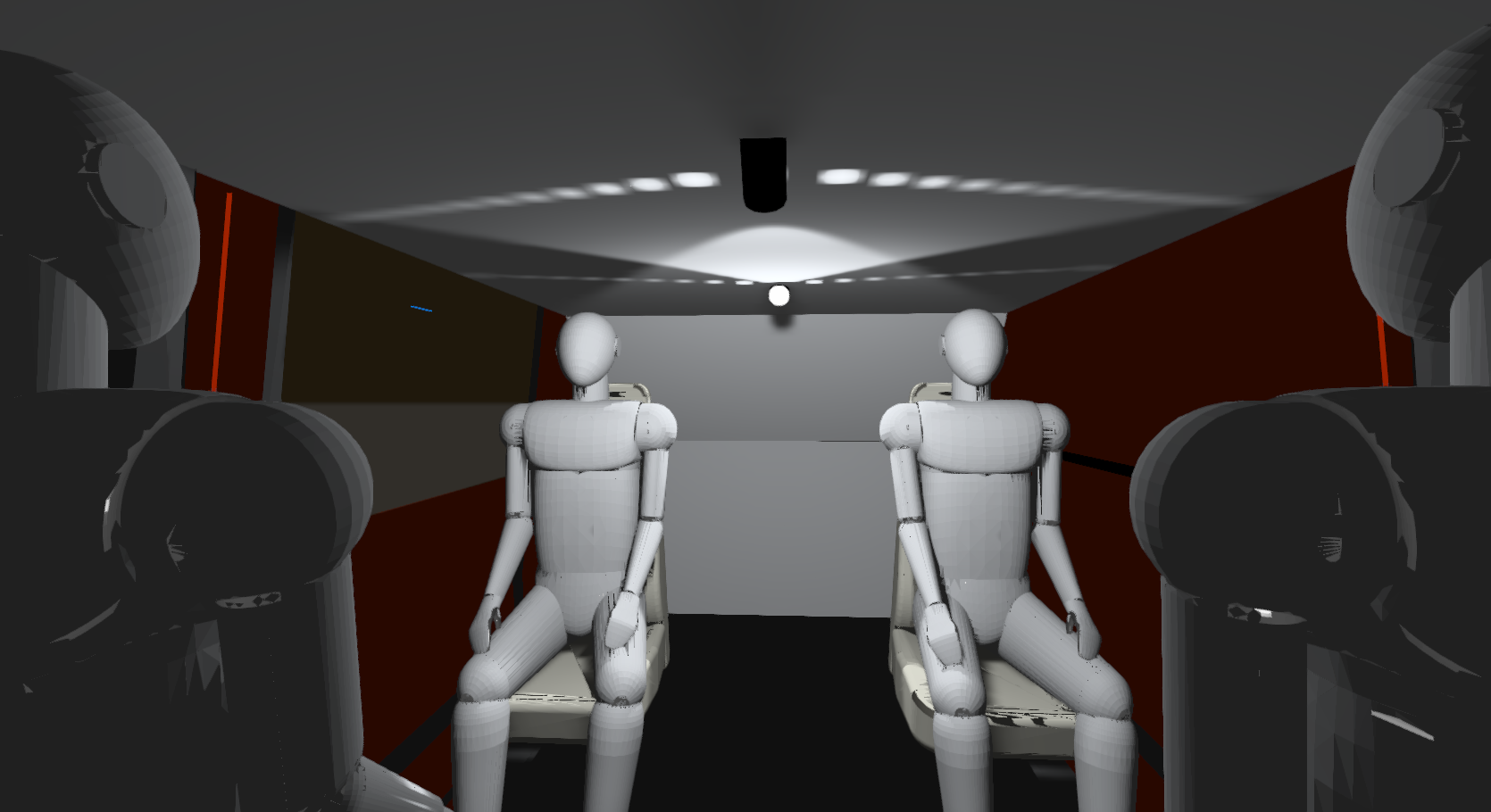}
\caption{Layout 6-1}
\label{fig:C6-1}
\end{subfigure}
~
\begin{subfigure}{0.3\linewidth}
\includegraphics [width=\linewidth]{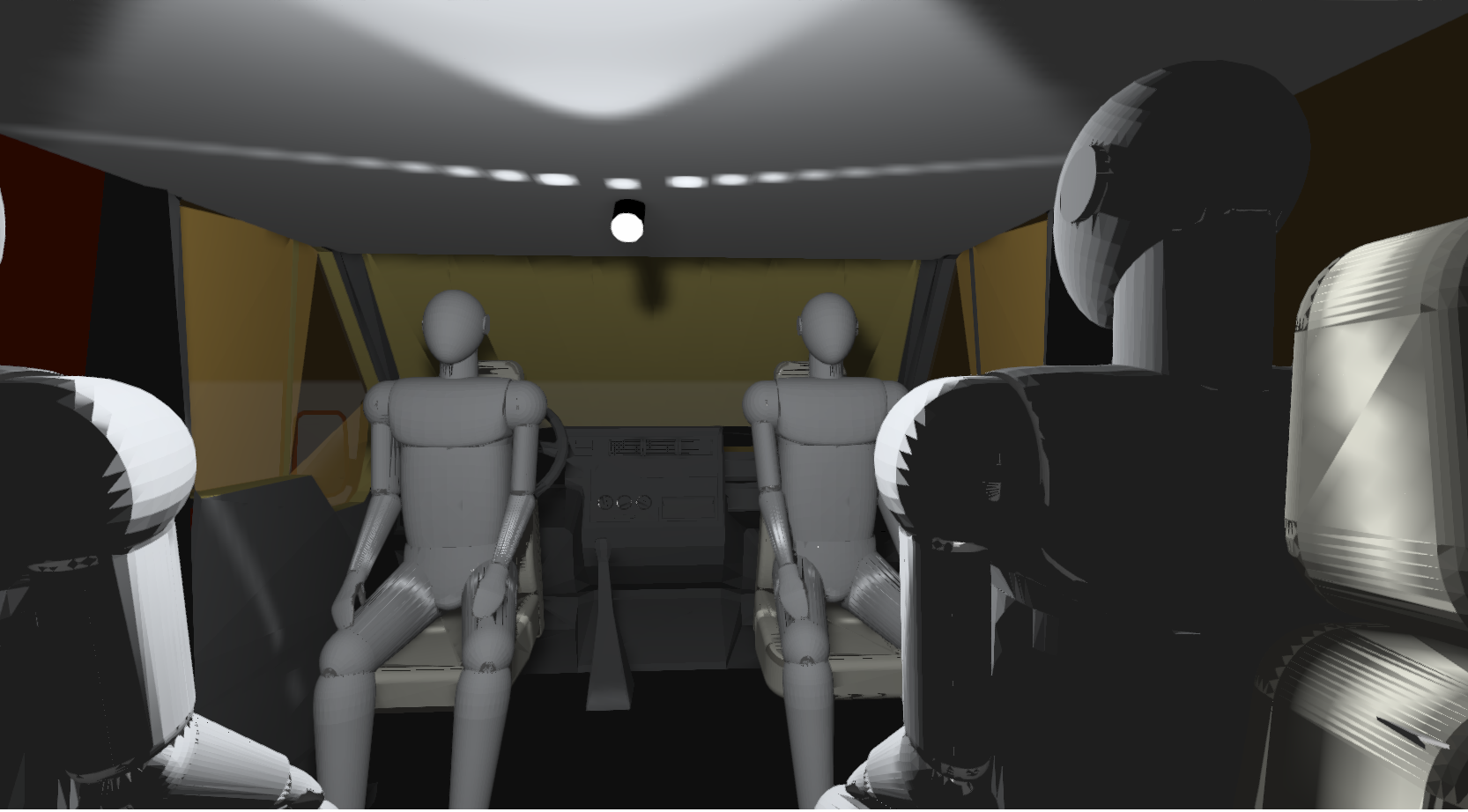}
\caption{Layout 6-2}
\label{fig:C6-2}
\end{subfigure}
\caption{Visualization of Coverage}

\label{fig:Result}
\end{figure}

As explained earlier, the camera is substituted with a light source (94-degree diagonal field of view). The light sources were positioned at the optimized camera locations (see Fig. \ref{fig:OptViz}) in the cabin, and the resulting visualization is given in Fig. \ref{fig:Result}. Here, the dark areas are the locations where the camera does not see. For example, in Fig. \ref{fig:C4}, the legs of both the occupants are not illuminated and thus look dark. This means that the cameras cannot see the legs of the occupants, while the upper body of the occupants are well illuminated, which infers that the cameras capture the whole upper body. Likewise, concluding from all the sub-figures of  Fig. \ref{fig:Result}, the cameras are able to capture all the markers on the upper-body of all the occupants in the six configurations, proving the concept of the experiment. 

\subsection{Experiment 2: Coverage of occupants in a highly configurable cabin}

\begin{figure}[h!]
\centering

    \includegraphics [scale=0.5] {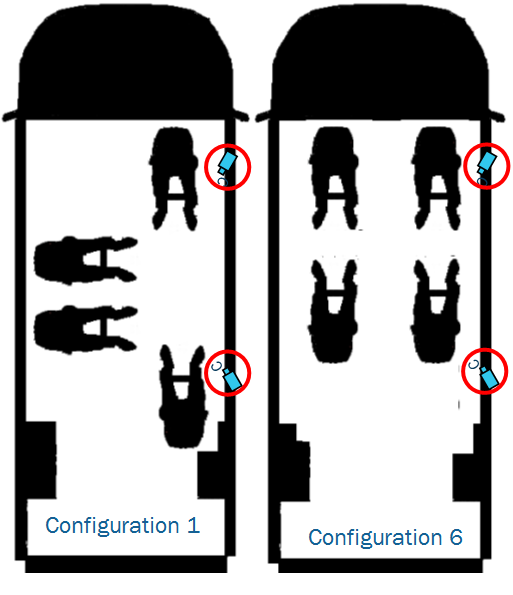}
\caption{Optimized Camera Positions for Configurations 1 and~6
}

\label{fig:Config16}
\end{figure}

\begin{figure}[h!]
\centering

    \includegraphics [scale=0.5] {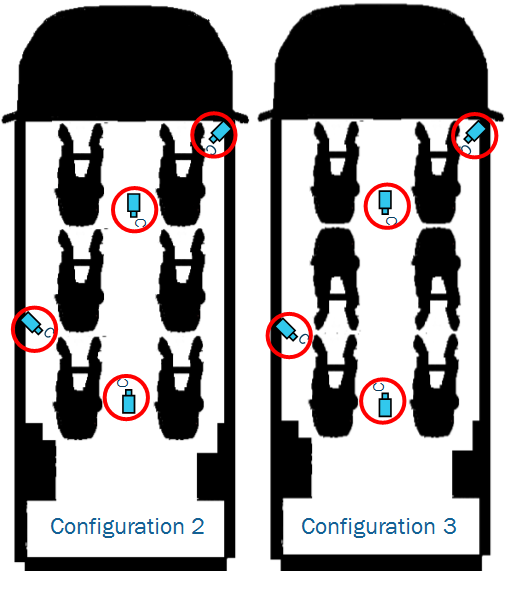}
\caption{Optimized Camera Positions for Configurations 2 and~3}

\label{fig:Config23}
\end{figure}

\begin{figure}[h!]
\centering

    \includegraphics [scale=0.5] {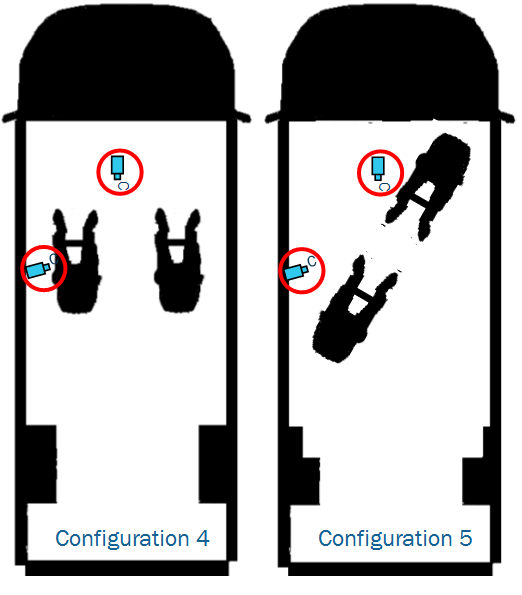}
\caption{Optimized Camera Positions for Configurations 4 and~5}

\label{fig:Config45}
\end{figure}

In this experiment, we assume that a cabin with certain number of seats can be reconfigured between different seat layouts. In Fig (\ref{fig:OptViz}), configuration 1 can be reconfigured to configuration 6 (Fig. (\ref{fig:Config16})), configuration 2 can be reconfigured to configuration 3 (Fig. (\ref{fig:Config23})), and configuration 4 can be reconfigured to configuration 5 (Fig. (\ref{fig:Config45})). Then assuming that the cameras will be fixed, the optimization goal is to find the camera numbers and locations that maximize the occupant coverage towards the combinations of configurations.

The results of the above mentioned optimized position are provided in Fig. (\ref{fig:Config16}), Fig. (\ref{fig:Config23}), and Fig. (\ref{fig:Config45}), respectively. When there are four seats in the cabin, two cameras are enough to cover all the four occupants in each of the two configurations (as shown in Fig. (\ref{fig:Config16}). When there are six seats, four cameras are needed to cover all the occupants' upper bodies (as shown in Fig. (\ref{fig:Config23})) as opposed to three cameras needed to cover each configuration. This is because the middle row in configuration 3 is facing rearward. If there are two seats, two cameras are still needed to cover all the body parts in a highly configurable cabin (as shown in Fig. (\ref{fig:Config45})). 

\subsection{Experiment 3: Maximum coverage of occupants given constraints on the number of cameras}

Although the experiments above show that as the number of camera increases, the cameras together can cover all the interested body parts from all occupants in different configurations. However, the total number of cameras is always limited in reality to reduce cost and system complexity. It is important to see the trade-offs between camera numbers and coverage rates. In this experiment, the optimization focuses on maximizing the coverage of occupant body parts, given a certain number of cameras. 

\begin{figure}[h!]
\centering

    \includegraphics [scale=0.5] {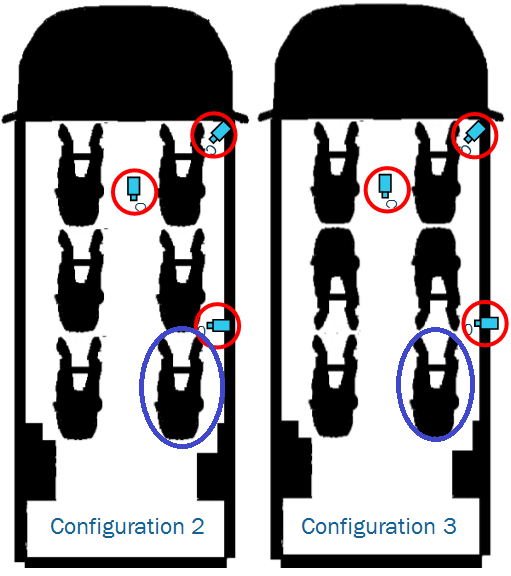}
\caption{Optimized Camera Positions for configurations 2 and 3 with three cameras}

\label{fig:Config23aspect3}
\end{figure}

The first trial assumes that there are only three cameras available for the six-seat highly configurable cabin (configuration 2 and 3). The optimization towards this combination with three cameras show that out of 72 body parts in total (across all occupants in all seats), 68 body parts can be covered with three cameras. The visualization of the optimization can be observed in Fig. \ref{fig:Config23aspect3}. Here, the uncovered points are the right shoulder and right hip of the occupants marked in blue circle.  

\begin{figure}[h!]
\centering

    \includegraphics [scale=0.5] {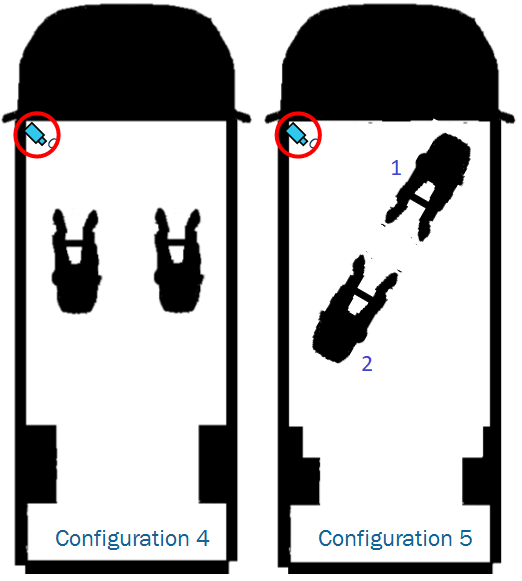}
\caption{Optimized Camera Positions for configurations 4 and 5 with three cameras}

\label{fig:Config45aspect3}
\end{figure}

Another trial focuses on using only one camera for the highly configurable cabin with two seats in total (configurations 4 and 5). The optimization results in a maximum coverage of 21 out of the 24 body parts (across all occupants in the two configurations). The visualization of the optimization is illustrated on Fig. \ref{fig:Config45aspect3}. 

The experiment results suggest that with limited total number of cameras, the body coverage is sacrificed. The trade-off between camera numbers and coverage rates needs to be investigated based on the sensing needs and algorithm development requirements. 

\section{Conclusion}


This paper peeks into the future camera-based sensing framework towards a cabin with highly configurable seat layouts in L3+ autonomous cars.  In particular, a proposed simulation and optimization process can estimate the number of cameras, their designs, and the coverage rates of occupant body parts in all possible seats. The research pictures the camera design concept for commonly-mentioned driver-less seat layouts, independently or combined, based on representative cabin dimensions and occupant sizes. 

We conducted three experiments using the proposed simulation and optimization method. Towards seat layouts with two, four, or six seats in total, the result shows that the sensing system needs one, two, or three cameras to cover all the occupants' upper bodies. When the cabin is highly configurable (meaning the seat layout can change from one to another completely), the sensing system may need more cameras at different poses and can still achieve full upper-body coverage. Not surprisingly, additional investigation shows that reducing the total number of cameras will sacrifice coverage rate to a certain degree. This finding suggests future development to focus on the trade-off (between camera numbers and coverage rates) towards specified sensing needs and system development requirements. 

\addtolength{\textheight}{-5cm}   




\bibliography{bib}

\end{document}